\documentclass[conference,10pt]{IEEEtran}
\usepackage{epsfig,rotating,setspace,latexsym,amsmath,epsf,amssymb,amsfonts,bm,subfigure,epstopdf,cite,authblk,bbm,mathtools,amsthm,algorithm,color,systeme,mathtools,amsmath,derivative,bigints}
\usepackage[utf8]{inputenc}
\usepackage[T1]{fontenc}
\usepackage[thinc]{esdiff}
\usepackage[noend]{algpseudocode}

\theoremstyle{plain}
\newtheorem{theorem}{Theorem}
\newtheorem{lemma}{Lemma}

\newenvironment{Proof}[1]{\medskip\par\noindent{\bf Proof:\,}\,#1}{{\mbox{\,$\blacksquare$}\par}}

\algrenewcommand\algorithmicforall{\textbf{foreach}}
\algrenewcommand\algorithmicindent{.8em}

\IEEEoverridecommandlockouts
\allowdisplaybreaks

\begin{document}

\title{To Re-transmit or Not to Re-transmit for Freshness}

\author{Subhankar Banerjee \qquad Sennur Ulukus \qquad Anthony Ephremides\\
\normalsize Department of Electrical and Computer Engineering\\
\normalsize University of Maryland, College Park, MD 20742\\
\normalsize \emph{sbanerje@umd.edu} \qquad \emph{ulukus@umd.edu} \qquad \emph{etony@umd.edu}}
	
\maketitle
\begin{abstract}
We consider a time slotted communication network with a base station (BS) and a user. At each time slot a fresh update packet arrives at the BS with probability $p>0$. When the BS transmits an update packet for the first time, it goes through with a success probability of $q_1$. In all subsequent re-transmissions, the packet goes through with a success probability of $q_2$ where $q_2>q_1$, due to the accumulation of observations at the receiver used to decode the packet. When the packet goes through the first time, the age of the user drops to 1, while when the packet goes through in subsequent transmissions, the age of the user drops to the age of the packet since its generation. Thus, when the BS is in the process of re-transmitting an old packet, if it receives a new packet, it has to decide whether to re-transmit the old packet with higher probability of successful transmission but resulting in higher age, or to transmit the new packet which will result in a lower age upon successful reception but this will happen with lower probability. In this paper, we provide an optimal algorithm to solve this problem.
\end{abstract}

\section{Introduction}
We consider a status update system, where a base station (BS) transmits update packets to a user over an erasure channel. Our goal is to minimize the age of information at the user by timely delivery of fresh update packets. Minimizing the age of information in various network settings has been extensively studied in the literature, e.g., \cite{kadota2019minimizing, moltafet2019closed, najm2018status, bedewy2019minimizing, costa2016age, yates2018age, moltafet2020age, 9478783, kaul2012status, arafa2018online, arafa2021timely}. The problems these papers solve can be broadly categorized into two groups: the first group of problems is to find closed-form expressions for the average or peak age under various network statistics, and the second group of problems is to find optimal scheduling policies for various network settings.

Different than all the aforementioned works, we consider the possibility of re-transmissions with higher probability of successful transmission. We assume that if the BS transmits a fresh update packet, then the user successfully receives it with probability $q_{1}$, and upon the failure of transmission for the first time, the BS can re-transmit the packet with higher probability of successful transmission $q_{2}$. This assumption is valid because the user can use a hybrid automatic repeat request (HARQ) protocol to accumulate previous observations whereby increasing the probability of successful reception. 

Minimizing the age of information by introducing coding and HARQ policies to combat the erasure channel for various network settings has been considered extensively in the literature. These works can be broadly classified in two categories: finding the optimal code lengths to minimize the age of information for various network settings \cite{najm2022optimal, sac2018age, chen2019benefits, you2021age}, and finding the optimal transmission policies for HARQ techniques under various network settings \cite{najm2017status, yates2017timely, arafa2019timely, ceran2021reinforcement, ceran2019average, feng2019age, farazi2020average, wang2020age, arafa2020timely, arafa2021sample, li2022age, baknina2018coded}.

In this paper, we answer the following question: What is the optimal action for the BS when it receives a fresh update packet while an old update packet is being served; should the BS preempt the old packet and transmit the new packet with lower probability of successful transmission $q_{1}$, or should it discard the new packet and continue to transmit the old packet with higher probability of successful transmission $q_{2}$. To find the optimal transmission policy, the above mentioned works  \cite{najm2017status, yates2017timely, arafa2019timely, ceran2021reinforcement, ceran2019average, feng2019age, farazi2020average, wang2020age, arafa2020timely, arafa2021sample, li2022age, baknina2018coded} utilize the knowledge of the HARQ policy being used in the system. In this paper, we are agnostic to the underlying HARQ policy employed by the system, which is fundamentally different from these works, and needs a different analysis to find the structure of the optimal policy.\footnote{For mathematical tractability, we assume that the successful transmission probability $q_2$ remains the same for every re-transmission. A more general version of the problem, where the probability of successful transmission depends on the number of re-transmissions, can be an interesting future work.} 

In this paper, we show that this problem has an optimal stationary deterministic policy. We find the structure of this optimal policy: Let $v_{1}$ denote the age at the user and $v_{2}$ denote the age of the old packet at the BS. Let the BS receive a new update packet. Then: We show that, if the optimal action for the BS is to transmit the new packet by preempting the old packet, then for all cases when the age of the user is $v_{1}$ and the age of the old packet at the BS is $v_{2}+x$ with $x>0$,  the optimal action is also to transmit the new packet by preempting the old one. On the other hand, if the optimal action for the BS is to discard the new packet and keep re-transmitting the old packet, then for all cases when the age at the user is $v_{1}+x$ with $x>0$, the age of the old packet at the BS is $v_{2}$, the optimal action is also to discard the new packet and keep re-transmitting the old one. Finally, leveraging these structural properties, we propose a relative value iteration algorithm to solve the problem optimally.

\section{System Model and Problem Formulation}

We consider a slotted wireless communication system, with one BS and one user with Geo/Geo/1/1 queue. The last $1$ in the Kendall's notation denotes that only $1$ update packet is kept in the system. At each time slot the BS receives a new update packet with probability $p>0$. If at time slot $t$, the queue is empty and the BS receives an update packet, then the BS transmits this update packet with probability of successful transmission of $q_{1}$ to the user. If the user does not successfully receive this update packet at time slot $t$, and if the BS does not receive any new update packet at time slot $t+1$, then the BS re-transmits this old update packet with probability of successful transmission $q_{2}$ to the user, where $q_{2}>q_{1}$.

The process of re-transmission continues till the BS receives a new update packet or the old update packet which was generated at time $t$ gets received by the user. If the latter case occurs at time slot $t+k$, then the age of the user drops down to $k$. However, if the former case occurs at time slot $t+k$, then the BS has to decide whether to keep transmitting the old update packet generated at $t$ with probability $q_{2}$ or to transmit the new update packet with probability $q_{1}$. At time slot $t+k$, if the BS chooses to transmit the new update packet,  the age of the user drops down to $1$ with probability $q_{1}$, otherwise, the age of the user drops down to $k$ with probability $q_{2}$. 

Consider an algorithm $\pi$, which decides at time slot $t$ whether the BS should re-transmit an old update packet or to transmit a newly available update packet, to the user. The action corresponding to the algorithm $\pi$, at time slot $t$ is defined as $\pi(t)\in\{0,1\}$, where $0$ means ``re-transmit the old packet'' and $1$ means ``transmit the new packet''. We call such an algorithm a switching algorithm. We define the set of all possible causal switching algorithms as ${\Pi}$. The age of the user at time $t$ is defined as $v(t) = t- u(t)$, where $u(t)$ is the generation time of the freshest update packet available to the user till time $t$. We solve the following optimization problem,
\begin{align}\label{eq:formu}
    \inf_{\pi\in{\Pi}} \limsup_{T\rightarrow\infty} \frac{1}{T} \sum_{t=0}^{T-1} \mathbb{E}_{\pi}[v(t)]
\end{align}

\section{Optimal Policy}

In this section, we formulate the problem in (\ref{eq:formu}) as an MDP and solve the MDP by means of value iteration and leveraging the structure of the optimal policy. First, we define the components of the MDP for this problem.

\textit{State:} The state of this MDP problem is a tuple with three elements $(v_{1}, v_{2}, b )$ where $v_{1}$ is the age at the user, $v_{2}$ is the age of the packet in the BS queue, and $b$ is an indicator variable, where $b=1$ implies that the BS has a fresh update packet to transmit and $0$ implies that it does not. We define $\mathcal{S}$ to be the set of all possible states. Note that the age of the system can possibly be infinite, thus, $\mathcal{S}$ is a countably infinite set. At time $t$, we denote the state of the system as $s(t)$. For example, if at time $t$, the last time the user successfully received an update packet was at time $t'$, and after time $t'$ the BS again received an update packet at time $t''$, and from $t''$ to $t$ the BS kept re-transmitting that packet where $t'<t''<t$, and also  that at time $t$, the BS has received another fresh update packet, then, $s(t) = (t-t', t-t'', 1)$. If there is no update packet in the queue, then we denote the state as $(v_{1},\infty,b)$.

\textit{Action space:} We define the action of the BS at time $t$ as $a({t})$, $a({t})\in{\{0,1,2\}}$, where $a({t})=1$ denotes that at time $t$, the BS transmits a freshly received update packet, $a({t})=0$ denotes that the BS re-transmits the old packet in the queue and discards the freshly received update packet (if any), and $a(t)=2$ denotes that the BS stays idle. Note that, if the state of the system is $(v_{1},v_{2},0)$, i.e., that there is an old packet in the queue and no new packet arrives, then the BS always chooses action $0$, i.e., it re-transmits the packet in the queue. Note also that, if the state of the system is $(v_{1}, \infty, 1)$, i.e., that there is no old packet in the queue, and a new packet arrives, then the BS always chooses action $1$, i.e., it transmits the newly arriving packet. Note further that, if the state of the system is $(v_{1}, \infty, 0)$, i.e., that there is no old packet in the queue and no new packet arrives, then the BS always chooses action $2$, i.e., it remains idle. 

\textit{Transition probabilities:}
We denote the transition probability from state $\hat{s}$ to state $\tilde{s}$ under action $a$ as $P_{a}(\hat{s},\tilde{s})$. Then, for the following states, 
\begin{align}
&s\!=\!(v_{1}, v_{2}, 0), \ s_{1}\!=\! (v_{1}\!+\!1, v_{2}\!+\!1, 1 ), \ s_{2} \!=\! (v_{1}\!+\!1, v_{2}\!+\!1, 0) \nonumber \\ 
&s_{3} \!=\! (v_{2}\!+\!1, \infty, 0), \ s_{4} \!=\! (v_{2}\!+\!1,\infty,1) 
\end{align}
we have the transition probabilities,
\begin{align}\label{trpr:3}
P_{0}(s,s_{1}) &=  p (1-q_{2}) \nonumber\\
P_{0}(s,s_{2}) &=  (1-p) (1-q_{2}) \nonumber\\
P_{0}(s,s_{3}) &=  (1-p) q_{2} \nonumber\\  
P_{0}(s,s_{4}) &=  p q_{2} 
\end{align}
Similarly, for the following states, 
\begin{align}
&s' \!=\! (v_{1}, v_{2}, 1), \ s_{1}' \!=\! (v_{1}\!+\!1, v_{2}\!+\!1, 1), \ s_{2}'\!=\! (v_{1}\!+\!1,v_{2}\!+\!1,0) \nonumber\\ 
&s_{3}' = (v_{2}+1,\infty,0), \ s_{4}' = (v_{2}+1, \infty,1), \ s_{5}'= (v_{1}+1,1,0) \nonumber\\
&s_{6}' = (v_{1}+1,1,1), \ s_{7}'=(1,\infty,0), \ s_{8}' = (1,\infty,1) 
\end{align}
we have the transition probabilities,
\begin{align}\label{trpr:5}
P_{0}(s',s_{1}') &=p (1-q_{2}) \nonumber \\
P_{0}(s',s_{2}') &=(1-p) (1-q_{2}) \nonumber \\
P_{0}(s',s_{3}') &= (1-p) q_{2} \nonumber \\  
P_{0}(s',s_{4}') &= p q_{2} \nonumber \\ 
P_{1}(s',s_{5}') &= (1-p) (1-q_{1}) \nonumber \\ 
P_{1}(s',s_{6}') &= p (1-q_{1}) \nonumber \\ 
P_{1}(s',s_{7}') &= (1-p) q_{1} \nonumber \\ 
P_{1}(s',s_{8}') &= p q_{1} 
\end{align}
Further, for the following states, 
\begin{align}
\bar{s}\!=\! (v_{1},\infty,0), \ \bar{s}_{1} \!=\! (v_{1}\!+\!1, \infty , 0), \ \bar{s}_{2} \!=\! (v_{1}\!+\!1, \infty , 1)
\end{align}
we have the transition probabilities,
\begin{align}\label{trpr:7}
P_{2}(\bar{s}, \bar{s}_{1}) &=  (1-p) \nonumber \\
P_{2}(\bar{s}, \bar{s}_{2}) &=  p
\end{align}
Finally, all other transition probabilities are zero.

\textit{Cost:}
If the state of the system is $\hat{s}$ and if the action taken by the BS is $a$, then we define the cost of the system as $C(\hat{s},a)$. In this paper, the cost of the system is the age at the user,
\begin{align}\label{eq:4}
    C(\hat{s},a) = \sum_{\tilde{s}\in{\mathcal{S}}} P_{a}(\hat{s},\tilde{s}) \tilde{v}
\end{align}
where $\tilde{v}$ is the first component of $\tilde{s}$, i.e., the age at the user. We call the MDP with the above mentioned state space, action space, transition probabilities and cost as $\Delta$. For a policy $\pi$, we define the average cost for $\Delta$ starting from state $\tilde{s}$, as
\begin{align}
    J^{\pi}(\tilde{s}) =  \limsup_{T\rightarrow \infty} \frac{1}{T} \sum_{t=0}^{T-1} \mathbb{E}_{\pi}\left[ C(s({t}),a({t}))|s({0})=\tilde{s}\right]
\end{align}
We define the optimal average cost starting from state $\tilde{s}$ as,
\begin{align}
    J^{*}(\tilde{s}) = \inf_{\pi\in\Pi} J^{\pi}(\tilde{s})
\end{align}
  Thus, (\ref{eq:formu}) becomes,
\begin{align}\label{eq:5}
   \!\!\!J^{*}(\tilde{s}) = \inf_{\pi\in{\Pi}} \limsup_{T\rightarrow \infty} \frac{1}{T} \sum_{t=0}^{T-1} \mathbb{E}_{\pi}\left[ C(s(t),a(t))|s(0)=\tilde{s}\right]
\end{align}

Next, we introduce the corresponding discounted MDP, which we use to solve the original MDP in (\ref{eq:5}). Consider a policy $\pi$ and  $\alpha\in{(0,1)}$, then for state $\tilde{s}\in{\mathcal{S}}$, we define the discounted cost as $V_{\pi,\alpha}(\tilde{s})$,
\begin{align}
    V_{\pi,\alpha}(\tilde{s}) = \mathbb{E}_{\pi} \left[\sum_{t=0}^{\infty} \alpha^{t} C(s({t}),a({t})) \Big| s_{0}=\tilde{s} \right]
\end{align}
We define the optimized $V_{\pi,\alpha}(\tilde{s})$,
\begin{align}
    V_{\alpha}(\tilde{s}) = \min_{\pi\in{\Pi}} V_{\pi,\alpha}(\tilde{s})
\end{align}
which is also known as the value function for state $\tilde{s}$. 

Next, we present Lemma~\ref{lemma:1} and Lemma~\ref{lemma:2}, which are well-known in the MDP literature, see \cite{puterman2014markov}, which we will use to find the structure of the optimal policy. We define $V_{\alpha}(\tilde{s};a)$ as,
\begin{align}
    V_{\alpha}(\tilde{s};a) = C(\tilde{s},a) + \alpha \sum_{\hat{s}\in{\mathcal{S}}} P_{a}(\tilde{s},\hat{s}) V_{\alpha}(\hat{s})
\end{align} 
\begin{lemma}\label{lemma:1}
For every $\tilde{s}\in\mathcal{S}$ and for every $\alpha\in{(0,1)}$, if $V_{\alpha}(\tilde{s})$ is finite, then $V_{\alpha}(\tilde{s})$ satisfies the following equation,
\begin{align}\label{eq:8}
    V_{\alpha}(\tilde{s}) = \min_{a\in{\{0,1,2\}}} V_{\alpha}(\tilde{s};a) 
\end{align}
\end{lemma}

In the MDP literature, (\ref{eq:8}) is also known as the discounted optimality equation. We define $\forall{\tilde{s}}\in{\mathcal{S}}$, $V_{\alpha,0}(\tilde{s}) =0$, and 
\begin{align}
    V_{\alpha,n}(\tilde{s};a) = C(\tilde{s},a) + \alpha \sum_{\hat{s}\in{\mathcal{S}}} P_{a}(\tilde{s},\hat{s}) V_{\alpha,n-1}(\hat{s})
\end{align}
and
\begin{align}
V_{\alpha,n}(\tilde{s}) = \min_{a\in{\{0,1,2\}}} V_{\alpha,n}(\tilde{s};a)
\end{align}

\begin{lemma}\label{lemma:2}
For all $\alpha \in{(0,1)}$ and $\tilde{s}\in\mathcal{S}$, $V_{\alpha,n}(\tilde{s})\rightarrow V_{\alpha}(\tilde{s})$.
\end{lemma}

We say that $f$ is a stationary deterministic policy for $\Delta$, if it always chooses the same action for a given state $\tilde{s}$ independent of the past decisions, $f(\tilde{s})$ denotes the decision for the state $\tilde{s}$. In the next theorem, we show the existence of a stationary deterministic policy which is optimal for $\Delta$. From \cite[Thm.]{sennott1989average}, we say that under some technical assumptions namely, $V_{\alpha}(\tilde{s})$ is finite, for all $\tilde{s}\in{\mathcal{S}}$, (\ref{eq:th1:18}), (\ref{eq:th1:19}) and (\ref{eq:th1:20}), there exists a stationary deterministic policy which is optimal for the MDP $\Delta$. We use \cite[Prop.~5]{sennott1989average} to show these technical assumptions, namely we find a stationary deterministic policy which induces an ergodic Markov chain and under this policy the expected traveling cost from any state to a fixed state is finite. 

\begin{theorem}\label{th:1}
There exists a stationary deterministic policy which minimizes
\begin{align}
\limsup_{T\rightarrow \infty} \frac{1}{T} \sum_{t=0}^{T-1} \mathbb{E}_{\pi}\left[ C(s({t}),a({t}))|s({0})=\tilde{s}\right].
\end{align}
\end{theorem}

\begin{Proof}
Consider a stationary policy $f$, for which the BS only transmits when it receives a fresh update packet, otherwise it stays idle. It is easy to see that the policy $f$ induces a Markov chain $M$ on the state space $\mathcal{S}$. Note that $M$ is an irreducible and aperiodic Markov chain. Now, we show that $M$ is also positive recurrent. From \cite[Thm.~1.27]{serfozo2009basics}, we know that for an irreducible Markov chain, if one state is positive recurrent, then all of the states are positive recurrent. Now, we show that the state $s_{8}'$ is positive recurrent. We define $\tau_{\tilde{s},\tilde{s}}$ to be the time needed to revisit state $\tilde{s}$ starting from the same state $\tilde{s}$, for the first time, formally $\tau_{\tilde{s},\tilde{s}} = \inf_{t\geq 1} \{t: s(t)=\tilde{s}, s(0)=\tilde{s}\}$,
\begin{align}\label{ne:9}
     \!\!\!  \mathbb{E}[\tau_{s_{8}',s_{8}'}] \!=\! p q_{1} + (1-pq_{1})\sum_{i=2}^{\infty} i p^{2} q_{1} (1- p^{2} q_{1})^{i-2} <  \infty \!
\end{align}
    
Now, we define $H_{\hat{s},\tilde{s}}$ to be the cost that $f$ suffers for visiting state $\tilde{s}$ for the first time starting from state $\hat{s}$. Then, $H_{\hat{s},\tilde{s}}$ is a random variable. For example, if $\hat{s}=(2,1,1)$ and $\tilde{s} = (1,\infty,1)$, for policy $f$, one trajectory is $(2,1,1)\rightarrow (3,1,1)\rightarrow (1,\infty,1)$ and the realization of $H_{\hat{s},\tilde{s}}$ corresponding to this trajectory is $C((3,1,1),1)$. Note that, if $\hat{s}=(\hat{v}_{1}, \hat{v}_{2}, b)$, then $C(\hat{s},1) \leq \hat{v}_{1}+1$. Now,
\begin{align}\label{ne:10}
        \mathbb{E}[H_{\hat{s},{s_{8}'}}] \!\leq \! (1\!-\!p q_{1}) p^{2} q_{1} \!\sum_{i=1}^{\infty} (1\!-\! p^{2} q_{1})^{i-1} \!\sum_{j=1}^{i} (v_{1} \!+\! j\!+\!1)  
        \!<\!   \infty
\end{align}

Based on (\ref{ne:9}) and (\ref{ne:10}), from \cite[Prop.~5]{sennott1989average}, for $0<\alpha<1$ and $\tilde{s}\in{\mathcal{S}}$, $V_{\alpha}(\tilde{s})$ is finite and for every state $\tilde{s}$, and that there exists a non-negative real number $M_{\tilde{s}}$, such that
\begin{align}\label{eq:th1:18}
    V_{\alpha}(\tilde{s}) - V_{\alpha}(s_{8}') \leq M_{\tilde{s}}
\end{align}
Moreover, $\tilde{s}\in{\mathcal{S}}$, there exists an action $a_{\tilde{s}}$ such that,
\begin{align}\label{eq:th1:19}
     \sum_{\hat{s}\in{\mathcal{S}}} P_{a_{\tilde{s}}}(\tilde{s},\hat{s}) M_{\hat{s}} < \infty
\end{align}
As $V_{\alpha}(\tilde{s}) < \infty$ for all $\tilde{s}$, we define $V_{\alpha}(s_{8}') = L$. Thus, 
\begin{align}\label{eq:th1:20}
     V_{\alpha}(\tilde{s}) - V_{\alpha}(s_{8}') \geq - L, \quad \tilde{s}\in{\mathcal{S}}
\end{align}
Thus, from \cite[Thm.]{sennott1989average}, we say that there exists a stationary deterministic policy which is optimal for the MDP $\Delta$. 
\end{Proof}

Next, we prove the monotonicity of the value function. 

\begin{lemma}\label{lemma:3}
For any non-negative real valued $x$, $y$, $v_{1}$ and $v_{2}$,
\begin{align}
     V_{\alpha}((v_{1}+x, v_{2}+y,b)) \geq V_{\alpha}((v_{1}, v_{2}, b)), \quad b\in{\{0,1\}}
\end{align}
Further,
\begin{align}
         V_{\alpha}((v_{1},\infty, b)) \leq V_{\alpha}((v_{1}+x,\infty,b))
\end{align}
\end{lemma}

\begin{Proof}
To be consistent with the state space $\mathcal{S}$, $v_{2}+y\leq v_{1}+x$. We prove the first statement of this lemma by induction. Note that, for the first statement we do not consider the action $a=2$, as the BS always has an update packet to transmit. Recall that, $s' = (v_{1}, v_{2}, 1)$ and we define $\tilde{s}_{1}= (v_{1}+x, v_{2}+y, 1)$. Then,
\begin{align}
    V_{\alpha,1}(s';1) = & C(s',1) = q_{1} + (1-q_{1})(v_{1} +1) \label{eq:9} \\
    V_{\alpha,1} (\tilde{s}_{1}; 1) = & C(\tilde{s}_{1},1) =  q_{1}+ (1-q_{1}) (v_{1}+x+1) \label{eq:10}
\end{align}
Similarly,     
\begin{align}
       \!\!\!\!V_{\alpha,1}(s' ; 0) \!=\! &C(s', 0) \!=\! q_{2} (v_{2}\!+\!1) \!+\! (1\!-\! q_{2}) (v_{1} \!+\!1) \label{eq:11} \\
       \!\!\!\!V_{\alpha,1} (\tilde{s}_{1} ; 0) \!= \! &C(\tilde{s}_{1},0) \!=\!q_{2} (v_{2}\!+\!y\!+\!1) \!+\! (1\!-\!q_{2}) (v_{1}\!+\!x\!+\!1) \!\!\label{eq:12}
\end{align}
From (\ref{eq:9})-(\ref{eq:12}), we note $V_{\alpha,1}(s';a)\leq V_{\alpha,1} (\tilde{s}_{1};a)$, $a\in{\{0,1\}}$. Thus, $V_{\alpha,1}(s') \leq V_{\alpha,1}(\tilde{s}_{1})$. Now, assume that the statement is true for $n-1$, then we show that the statement is true for $n$,
\begin{align}\label{eq:13}
    V_{\alpha,n} (s';0) = C(s',0) + \sum_{\hat{s}\in{\mathcal{S}}} P_{0}(s',\hat{s}) V_{\alpha,n-1}(\hat{s})
\end{align}
and similarly, 
\begin{align}\label{eq:14}
        V_{\alpha,n}(\tilde{s}_{1};0) = C(\tilde{s}_{1},0) + \sum_{\tilde{s}\in{\mathcal{S}}} P_{0}(\tilde{s}_{1},\tilde{s}) V_{\alpha,n-1}(\tilde{s})
\end{align}
Now, we observe that, for a given state, $\hat{s}\in{\mathcal{S}}$, such that $P_{0}(s',\hat{s})>0$, there exists another state $\tilde{s}\in{\mathcal{S}}$, such that $\tilde{s}\geq \hat{s}$ and $P_{0}(s',\hat{s}) = P_{0}(\tilde{s}_{1},\tilde{s})$. Thus, following the induction step we say that,
\begin{align}
    \!\! \sum_{\hat{s}\in{\mathcal{S}}} P_{0}(s',\hat{s}) V_{\alpha,n-1}(\hat{s})\leq \sum_{\tilde{s}\in{\mathcal{S}}} P_{0}(\tilde{s}_{1},\tilde{s}) V_{\alpha,n-1}(\tilde{s})
\end{align}
and from (\ref{eq:11})  and (\ref{eq:12}), we know that $C(s',0)\leq C(\tilde{s}_{1},0)$. Thus, combining these two facts, we have
\begin{align}
    V_{\alpha,n}(s';0)\leq V_{\alpha,n}(\tilde{s}_{1};0)
\end{align}
Similarly, we have
\begin{align}
    V_{\alpha,n}(s';1)\leq V_{\alpha,n}(\tilde{s}_{1};1)
\end{align}
Combining these, we obtain
\begin{align}
    V_{\alpha,n}(s') & = \min_{a\in{\{0,1\}}} \{V_{\alpha,n}(s';a)\} \\ 
    & \leq \min_{a\in{\{0,1\}}}\{V_{\alpha,n}(\tilde{s}_{1};a)\} \\
    & = V_{\alpha,n}(\tilde{s}_{1})
\end{align}
Thus, the first part of this lemma follows from Lemma~\ref{lemma:2}. Next, when the state is of the form $(v_{1}+x,\infty,b)$, then depending on $b$, the BS has only one choice, namely, if $b=1$ the BS always chooses action $1$, and if $b=0$ the BS stays idle, i.e., it chooses action $a=2$. Now, from Lemma~\ref{lemma:1} and the first part of this lemma, we have the second part of this lemma.
\end{Proof}

Now, we investigate a structure of the optimal policy. 
\begin{lemma}\label{lemma:4}
If the optimal policy for the state $s'=(v_{1}, v_{2}, 1)$ is to choose action $1$, then for $0\leq x\leq v_{1} -v_{2}$, the optimal policy for the state $(v_{1}, v_{2}+x, 1)$ is also to choose action $1$.
\end{lemma}
\begin{Proof}
We first prove this lemma for the $\alpha$-optimal policy. From the assumption of this lemma, we have
\begin{align} \label{nn:eq:21}
        V_{\alpha}(s'; 1) - V_{\alpha}((v_{1}, v_{2}, 1) ;0) \leq 0 
\end{align}
Expanding the left hand side of (\ref{nn:eq:21}) and using Lemma~\ref{lemma:3}, 
\begin{align}
        V_{\alpha}((v_{1}, v_{2}+x , 1);1) - V_{\alpha}((v_{1}, v_{2}+x,1);0) \leq 0
\end{align}
Thus, for $0<\alpha<1$ there exists an $\alpha$-optimal policy which has the structure given in the statement of the lemma. Now, consider a sequence $\{\alpha_{n}\}_{n=1}^{\infty}\subset(0,1)$, such that $\lim_{n\rightarrow\infty} \alpha_{n}=1$, then from \cite[Lem.~1]{sennott1989average}, there exists a sub-sequence $\{\beta_{n}\}_{n=1}^{\infty}$ and an optimal stationary policy $f$ such that $\lim_{n\rightarrow\infty} f_{\beta_{n}}(s) = f(s) $, $s\in{\mathcal{S}}$. Thus, the stationary policy $f$ also has the structure given in the lemma. 
\end{Proof}

The state space $\mathcal{S}$ is countably infinite, to prove Lemma~\ref{lemma:7} and to apply a value iteration algorithm, we need a finite state space. Thus, we consider an approximating sequence of MDPs, $\{\Delta_{N}\}_{N=1}^{\infty}$ to approximate $\Delta$. The
state space $\mathcal{S}_{N}$ associated with $\Delta_{N}$ has the states of the form similar to the states for $\mathcal{S}$, i.e., $(v_{1}, v_{2}, b)$, where $1\leq v_{1}\leq N$, $1\leq v_{2} \leq v_{1}$, $v_{2}=\infty$ and $b\in{\{0,1\}}$. Note that, $\mathcal{S}_{N}\uparrow \mathcal{S}$. The action space for $\Delta_{N}$ is the same as the action space of $\Delta$. To define the transition probabilities for $\Delta_{N}$, we first define a function $p_{\hat{s}}(\check{s},\tilde{s},N)$, note that, $\hat{s}=(\hat{v}_{1}, \hat{v}_{2}, \hat{b})$, $\check{s} = (\check{v}_{1}, \check{v}_{2}, \check{b})$ and $\tilde{s} = (\tilde{v}_{1}, \tilde{v}_{2}, \tilde{b})$.
\begin{align}
p_{\hat{s}}(\check{s},\tilde{s},a,N) = \begin{cases}
1, & \forall{\tilde{s}\in{\mathcal{S}\backslash\mathcal{S}_{N}}}, \hat{v}_{1}=\check{v}_{1}, \check{s}\in{\mathcal{S}_{N}}\\ 
& \hat{v}_{2} = \min\{{\tilde{v}_{2},\check{v}_{2}}\}, \hat{b} = \tilde{b}\\
0, &\text{otherwise}
\end{cases}
\end{align}
Now, we define the transition probabilities for $\Delta_{N}$, i.e., $P_{a}(\check{s},\hat{s};N)$, where $\check{s},\hat{s}\in{\mathcal{S}_{N}}$ and $a\in{\{0,1,2\}}$ as,
\begin{align}
     P_{a}(\check{s},\hat{s};N) = \sum_{\tilde{s}\in{\mathcal{S}\backslash\mathcal{S}_{N}}} P_{a}(\check{s},\tilde{s}) p_{\hat{s}}(\check{s},\tilde{s},a,N) 
\end{align}
If $1\leq \check{v}_{1}<N$, $1\leq \check{v}_{2}\leq \check{v}_{1}$ and $\check{b}\in{\{0,1\}}$, then
\begin{align}
    \hat{\mathcal{S}}=\{\hat{s}: P_{a}(\check{s},\hat{s})>0\} \subset \mathcal{S}_{N}
\end{align}
Thus, for this case, we define $P_{a}(\check{s},\hat{s};N) = P_{a}(\check{s},\hat{s})$. From \cite{sennott2009stochastic}, we say that,
\begin{align}
    \lim_{N\rightarrow{\infty}} P_{a}(\check{s},\hat{s};N) = P_{a}(\check{s},\hat{s})
\end{align}
Finally, the cost for the state action pair $(\check{s},a)$ is the same as the cost for the same state action pair $(\check{s},a)$ in $\Delta$,  $\check{s}\in{\mathcal{S}_{N}}$.

Let us define the $\alpha$-discounted optimal cost and the $\alpha$-discounted cost corresponding to a policy $\pi$ for the MDP $\Delta_{N}$, as $V_{\alpha}^{N}(\check{s})$ and $V_{\pi,\alpha}^{N}(\check{s})$ respectively, where $0<\alpha<1$ and $\check{s}\in{\mathcal{S}_{N}}$. In the next lemma, we show the convergence of a sequence of average cost optimal policies $\{\pi_{N}\}_{N=1}^{\infty}$ on $\{\Delta_{N}\}_{N=1}^{\infty}$ to an average cost optimal policy $\pi$ on $\Delta$. For the definition of convergence of a sequence of policies, see \cite{sennott2009stochastic}. This lemma will be useful to prove Theorem~\ref{th:2}.

To prove this lemma, according to \cite{sennott1997computing}, we first show that, for all $N\geq 1$, $0<\alpha<1$ and $\check{s}\in{\mathcal{S}_{N}}$,
\begin{align}
    V_{\alpha}^{N}(\check{s}) - V_{\alpha}^{N}(\tilde{s}_{2})\geq L
\end{align}
where $L\in{\mathbb{R}}$ and $\tilde{s}_{2}=(1,1,1)$. Then, we show that $J^{*}(\hat{s})<\infty$, $\hat{s}\in{\mathcal{S}}$. After which, we show that
\begin{align}
    \sum_{\hat{s}\in{\mathcal{S}_{N}}} p_{\hat{s}}(\check{s},\tilde{s},a,N) V_{\alpha,n}^{N}(\hat{s}) \leq V_{\alpha,n}^{N}(\tilde{s})
\end{align}
where $\check{s}\in{\mathcal{S}_{N}}$, $a\in{\{0,1,2\}}$ and $\tilde{s}\notin {\mathcal{S}_{N}}$.
Finally, we show that there exists a stationary policy $f$ under which,
$H_{\hat{s},\tilde{s}_{2}}$ and $\tau_{\hat{s},\tilde{s}_{2}}$ are finite for all $\hat{s}\in{\mathcal{S}}$, and
\begin{align}
    \sum_{\hat{s}\in{\mathcal{S}_{N}\backslash\tilde{s}_{2}}} p_{\hat{s}}(\check{s},\tilde{s},a,N) H_{\hat{s},\tilde{s}_{2}} \leq H_{\tilde{s},\tilde{s}_{2}}
\end{align}
where $\tilde{s}\notin \mathcal{S}_{N}$, $\check{s}\in{\mathcal{S}_{N}}$, $H_{\hat{s},\tilde{s}_{2}}$ and $\tau_{\hat{s},\tilde{s}_{2}}$ are the average traveling cost and the average travel time to the state $\tilde{s}_{2}$ from any state $\hat{s}$, for the first time, under the policy $f$.

\begin{lemma}\label{lemma:5}
     If $\pi$ is a limit point of a sequence of average cost optimal policies, $\{\pi_{N}\}_{N=1}^{\infty}$ for $\{\Delta_{N}\}_{N=1}^{\infty}$, then $\pi$ is an average cost optimal policy for $\Delta$.  
\end{lemma}

\begin{Proof}
 Let us consider a policy $\tilde{\pi}$, for which the BS remains idle for the entire time horizon. Note that, $V_{\tilde{\pi},\alpha}^{N}(\tilde{s}_{2})$ is an increasing function of $N$, thus the limit exists and we denote,
\begin{align}
    \lim_{N\rightarrow\infty}V_{\tilde{\pi},\alpha}^{N}(\tilde{s}_{2}) = V_{\tilde{\pi},\alpha}^{\infty}(\tilde{s}_{2})
\end{align}
Thus,
\begin{align}
    V_{\tilde{\pi},\alpha}^{\infty}(\tilde{s}_{2}) =& \sum_{t=0}^{\infty} \alpha^{t} C(s({t}),2) \\ 
    = &  \sum_{t=0}^{\infty} \alpha^{t} (t+2) \leq \frac{1}{1-\alpha} + \frac{1}{(1-\alpha)^{2}}
\end{align}
Thus, $V_{\alpha}^{N}(\check{s})-V_{\alpha}^{N}(\tilde{s}_{2}) \geq -\left(\frac{1}{1-\alpha} + \frac{1}{(1-\alpha)^{2}}\right)$, $\check{s}\in{\mathcal{S}_{N}}$, $N\geq1$ and $0<\alpha<1$.  

Consider the policy $f$ in the proof for Theorem~\ref{th:1}, i.e., the BS transmits only if it receives a fresh update packet and remains idle otherwise. Similar to (\ref{eq:formu}) and (\ref{eq:5}), for the policy $f$ and the initial state $\tilde{s}_{2}$, we say that,
\begin{align}\label{eq:pr21}
    J^{f}(\tilde{s}_{2}) = \limsup_{T\rightarrow \infty} \frac{1}{T} \sum_{t=0}^{T-1} \mathbb{E}_{f} [{v(t)}|s(0)=\tilde{s}_{2}]
\end{align}
Note that, the stochastic process in (\ref{eq:pr21}), $\{v(t)\}_{t=1}^{\infty}$ is a renewal process, we define the inter-renewal time as $\tau_{r}$. We claim that $\mathbb{E}[\tau_{r}]< \infty$. Thus, from the renewal reward theorem \cite{serfozo2009basics}, we say that $J^{f}(\tilde{s}_{2})<\infty$. Thus, $J^{*}(\tilde{s}_{2}) <\infty$. Note that, $f$ induces an uni-chain on the state space $\mathcal{S}$, i.e., the induced Markov chain has only one recurrent class plus some transient states. Thus, from \cite{bertsekas2012dynamic}, we claim that,
\begin{align}
    J^{*}(\check{s})= J^{*}(\tilde{s}_{2})<\infty, \quad \check{s}\in{\mathcal{S}}
\end{align}
If $p_{\hat{s}}(\check{s},\tilde{s},a,N)>0$ and $\tilde{v}_{2}\neq \infty$, then $\hat{s}$ is of the form $(N, \min{\{\check{v}_{2}}, \tilde{v}_{2}\}, \tilde{b})$, note that $\tilde{s}\geq \hat{s}$, thus from the proof of Lemma~\ref{lemma:3}, we say that,
\begin{align}
\sum_{\hat{s}\in{\mathcal{S}_{N}}} p_{\hat{s}}(\check{s},\tilde{s},a,N) V_{\alpha,n}^{N}(\hat{s}) \leq V_{\alpha,n}^{N}(\tilde{s})
\end{align}
for $n\geq 1$, $N\geq 1$, $\alpha\in{(0,1)}$, $\tilde{s}\notin{\mathcal{S}_{N}}$, $\check{s}\in{\mathcal{S}}_{N}$ and $a\in{\{0,1,2\}}$. We say that the similar holds true when $\tilde{v}_{2} = \infty$. 

Consider again the policy $f$. Under this policy, we can show that $H_{\check{s},\tilde{s}_{2}}$ and $\tau_{\check{s},\tilde{s}_{2}}$ are finite numbers. From the definition of $p_{\hat{s}}(\check{s},\tilde{s},a,N)$ and $H_{\tilde{s},\tilde{s}_{2}}$, we say that,
\begin{align}
    \sum_{\hat{s}\in{\mathcal{S}_{N}\backslash\tilde{s}_{2}}} p_{\hat{s}}(\check{s},\tilde{s},a,N) H_{\hat{s},\tilde{s}_{2}} \leq H_{\tilde{s},\tilde{s}_{2}}
\end{align}
for $n\geq 1$, $N\geq 1$, $\alpha\in{(0,1)}$, $\tilde{s}\notin{\mathcal{S}_{N}}$ and $\check{s}\in{\mathcal{S}}_{N}$. Combining all the above results, the statement of this lemma directly follows from \cite{sennott1997computing}.
\end{Proof}

Note that, $\mathcal{S}=\mathbb{N}^{2}\times \{0,1\}$. Now, we consider another MDP with the state space $\bar{\mathcal{S}} =\mathbb{R}^{2} \times\{0,1\} \cup s_{a}$. The cost of this new MDP for a state action pair $(\hat{s},a)$, is the same as (\ref{eq:4}), $\hat{s}\in{\bar{\mathcal{S
}}}$. The transition probabilities are defined as, (\ref{trpr:3}), (\ref{trpr:5}) and (\ref{trpr:7}). We call this new MDP as $\bar{\Delta}$. Let us define the discounted optimal cost for $\bar{\Delta}$ as $\bar{V}_{\alpha}(\check{s})$, $0<\alpha<1$, $\check{s}\in{\bar{\mathcal{S}}}$. In the next few lemmas, we study further structure of the value function, to get more structural results for the optimal policy.

\begin{lemma}\label{lemma:6}
 For a state $(v_{1}, v_{2}, b) \in{\bar{\mathcal{S}}}$, the value function $\bar{V}_{\alpha}((v_{1}, v_{2}, b))$ is a concave function of $v_{1}$, $v_{1}\geq v_{2}$ .
\end{lemma}
Lemma~\ref{lemma:6} can be proved by using mathematical induction. 

In a manner similar to constructing $\bar{\mathcal{S}}$ from $\mathcal{S}$, we construct $\bar{\mathcal{S}}_{N}$ from $\mathcal{S}_{N}$, i.e., $\bar{\mathcal{S}}_{N} = [0,N]^{2} \times\{0,1\}$. We define a sequence of MDPs $\{\bar{\Delta}_{N}\}_{N=1}^{\infty}$, from $\{\Delta_{N}\}_{N=1}^{\infty}$ in a similar manner we defined $\bar{\Delta}$ from $\Delta$. Let us define the discounted optimal cost corresponding to ${\bar{\Delta}}_{N}$ as $\bar{V}_{\alpha}^{N}(\check{s})$, $\check{s}\in{\bar{\mathcal{S}}_{N}}$.

In the next lemma, we show an interesting structure for the discounted optimal cost for $\Delta_{N}$, i.e., $V_{\alpha}^{N}(\check{s})$, $\check{s}\in{\mathcal{S}_{N}}$. Specifically, we show (\ref{lemm7:59}) and (\ref{lemm7:60}). To prove these, we first show that the left derivative of the discounted optimal cost for $\bar{\Delta}_{N}$ follows (\ref{eq:29}) and (\ref{eq:30}), then we integrate both sides of (\ref{eq:29}) and (\ref{eq:30}) with proper limits to get (\ref{lemm7:59}) and (\ref{lemm7:60}).

\begin{lemma}\label{lemma:7}
    For $y\geq 0$, if $v_{1}+y\leq N$ and $v_{2}\leq \bar{v}_{2} \leq v_{1}$, $\bar{v}_{2} \neq \infty$, then for $N\geq 1$, $b\in{\{0,1\}}$,
    \begin{align}\label{lemm7:59}
        {V}_{\alpha}^{N}((v_{1}&+y,v_{2},b))-{V}_{\alpha}^{N}((v_{1}, {v}_{2}, b )) \nonumber\\ & \leq {V}_{\alpha}^{N}((v_{1}+y, \bar{v}_{2},b)) - {V}_{\alpha}^{N}((v_{1},\bar{v}_{2},b))
    \end{align}
    Moreover,
    \begin{align}\label{lemm7:60}
       \!\!\!\! {V}_{\alpha}^{N}&((v_{1}+y, \bar{{v}}_{2},b)) - {V}_{\alpha}^{N}((v_{1},\bar{v}_{2},b))\nonumber\\&\leq \left(\frac{1\!-\!q_{1}}{1\!-\!q_{2}}\right) \!\Big({V}_{\alpha}^{N}\!((v_{1}+y,v_{2},b))-{V}_{\alpha}^{N}((v_{1}, {v}_{2}, b ))\Big) \!\!\!
    \end{align}
\end{lemma}

\begin{Proof}
Note that as $v_{2}$ and $\bar{v}_{2}$ cannot be infinity, i.e., there is always a packet to transmit for both the states $(v_{1}, v_{2},b)$ and $(v_{1}, \bar{v}_{2}, b)$, we omit the $a=2$ in the following. We first prove the statement of this lemma for $\bar{V}_{\alpha}^{N}(\check{s})$, $\check{s}\in{\bar{\mathcal{S}}_{N}}$. First, note that Lemma~\ref{lemma:6} is true for $\bar{V}_{\alpha}^{N}((v_{1}, v_{2}, b))$, thus $\bar{V}_{\alpha}^{N}((v_{1}, v_{2}, b))$ is concave in $v_{1}$. Thus, the left derivative of $\bar{V}_{\alpha}^{N}((v_{1},v_{2},b))$ with respect to $v_{1}$ exists for all $v_{1}\in{[1,N]}$ \cite{bertsekas2009convex}. Let us define the left derivative of $\bar{V}_{\alpha}^{N}((v_{1},v_{2}, b))$ with respect to $v_{1}$ as
\begin{align}
     \pdv{\bar{V}_{\alpha}^{N}((v_{1},\!v_{2},\!b))^{-}}{v_{1}}\! = \!\!\lim_{h\rightarrow 0^+} \!\frac{\bar{V}_{\alpha}^{N}((v_{1},\!v_{2},\!b)) \!-\! \bar{V}_{\alpha}^{N}((v_{1}\!-\!h,\!v_{2},\!b))}{h}
\end{align}

First, we show that for $1\leq v_{1}\leq N$,
\begin{align}
     \pdv{\bar{V}_{\alpha}^{N}((v_{1},v_{2},b))^{-}}{v_{1}} & \leq \pdv{\bar{V}_{\alpha}^{N}((v_{1},\bar{v}_{2},b))^{-}}{v_{1}} \label{eq:29} \\ 
     \pdv{\bar{V}_{\alpha}^{N}((v_{1},\bar{v}_{2},b))^{-}}{v_{1}} & \leq \left(\frac{1-q_{1}}{1-q_{2}}\right)\pdv{\bar{V}_{\alpha}^{N}((v_{1},{v}_{2},b))^{-}}{v_{1}} \label{eq:30}
\end{align}
We prove both (\ref{eq:29}) and (\ref{eq:30}) simultaneously by mathematical induction. We assume that if the optimal action for state $(v_{1}, v_{2}, 1)$ is $a=1$ $(a=0)$, then there exists $\delta>0$, such that the optimal action for state $(v_{1}',v_{2},1)$ is also $a=1$ $(a=0)$, where $|v_{1}'-v_{1}|\leq \delta$. Note that because of this assumption, the optimal action for state $(v_{1}, v_{2}, 1)$ is the same as the optimal action for state $(v_{1}-h, v_{2}, 1)$, for sufficiently small $h$. Also note that, for state $(v_{1}, v_{2}, 0)$ the optimal action is $a=0$. 

Note that Lemma~\ref{lemma:4} is also true for $\bar{\Delta}_{N}$, i.e., if the optimal action for  state $s'=(v_{1}, v_{2}, 1)$ is $a=1$, then the optimal action for state $\bar{s}_{3}=(v_{1}, \bar{v}_{2}, 1)$ is also $a=1$. Note that the above statement is true even for the inductive stages. Now, we assume that $b=1$ and for the induction step $n=1$, if the optimal action for state $s'$ is $a=1$, then from the above, the optimal action at $n=1$, for state $\bar{s}_{3}$ is also $1$. Thus,
\begin{align}\label{eq:31}
     \pdv{\bar{V}_{\alpha,1}^{N}(s')^{-}}{v_{1}} = (1- q_{1}) =  \pdv{\bar{V}_{\alpha,1}^{N}(\bar{s}_{3})^{-}}{v_{1}}
\end{align}
Now, consider the case when the optimal action at $n=1$ for state $\bar{s}_{3}$ is $a=0$, then again from Lemma~\ref{lemma:4}, the optimal action at $n=1$ for the state $s'$ is also $a=0$. Thus,
\begin{align}\label{eq:32}
     \pdv{\bar{V}_{\alpha,1}^{N}(s')^{-}}{v_{1}} = (1- q_{2}) =  \pdv{\bar{V}_{\alpha,1}^{N}(\bar{s}_{3})^{-}}{v_{1}}
\end{align}
Finally, consider the case when the optimal action at $n=1$ for state $\bar{s}_{3}$ is $a=1$, however the optimal action at $n=1$ for state $s'$ is $a=0$. Then,
\begin{align}\label{eq:33}
    \pdv{\bar{V}_{\alpha,1}^{N}(s')^{-}}{v_{1}} = (1- q_{2}) 
    < (1- q_{1}) = \pdv{\bar{V}_{\alpha,1}^{N}(\bar{s}_{3})^{-}}{v_{1}}                                        
\end{align}
Combining (\ref{eq:31}), (\ref{eq:32}) and (\ref{eq:33}), we say that,
\begin{align}\label{eq:34}
     \pdv{\bar{V}_{\alpha,1}^{N}(s')^{-}}{v_{1}} \leq  \pdv{\bar{V}_{\alpha,1}^{N}(\bar{s}_{3})^{-}}{v_{1}}
\end{align}
In a similar fashion, we show the following inequalities,
\begin{align}
     \pdv{\bar{V}_{\alpha,1}^{N}(\bar{s}_{3})^{-}}{v_{1}} & \leq \left(\frac{1-q_{1}} {1-q_{2}}\right)  \pdv{\bar{V}_{\alpha,1}^{N}(s')^{-}}{v_{1}} \label{eq:35} \\
     \pdv{\bar{V}_{\alpha,1}^{N}((v_{1},v_{2},0))^{-}}{v_{1}} & \leq  \pdv{\bar{V}_{\alpha,1}^{N}((v_{1},\bar{v}_{2},0))^{-}}{v_{1}} \label{eq:36} \\ 
     \pdv{\bar{V}_{\alpha,1}^{N}((v_{1},\bar{v}_{2},0))^{-}}{v_{1}} & \leq \left(\frac{1-q_{1}}{1-q_{2}}\right) \pdv{\bar{V}_{\alpha,1}^{N}((v_{1},v_{2},0))^{-}}{v_{1}} \label{eq:37}
\end{align}

Now, assume that all the inequalities are true for the inductive stage $n-1$. We first prove the inequalities for the induction stage $n$, for $v_{1}\leq N-1$. First, we consider the case when the optimal action for both states $s'$ and $\bar{s}_{3}$ at the induction stage $n$ is $a=0$. Then, for this case we write,
\begin{align}\label{eq:38}
    \bar{V}_{\alpha,n}^{N}(s') = & q_{2} (v_{2}+1) + (1-q_{2}) (v_{1}+1)\nonumber\\
    & + p q_{2} \bar{V}_{\alpha,n-1}^{N}((v_{2}+1, \infty,1)) \nonumber\\ 
    & + (1-p) q_{2} \bar{V}_{\alpha,n-1}^{N} ((v_{2}+1, \infty, 0)) \nonumber\\  
    &+ (1-p)(1-q_{2}) \bar{V}_{\alpha,n-1}^{N}((v_{1}+1, v_{2}+1, 0))  \nonumber\\ &+ p (1-q_{2}) \bar{V}_{\alpha,n-1}^{N}((v_{1}+1, v_{2}+1,1))
\end{align}
Similarly, 
\begin{align}\label{eq:39}
    \bar{V}_{\alpha,n}^{N}(\bar{s}_{3}) =& q_{2} (\bar{v}_{2}+1) + (1-q_{2}) (v_{1}+1)\nonumber\\
    & + p q_{2} \bar{V}_{\alpha,n-1}^{N}((\bar{v}_{2}+1, \infty,1)) \nonumber\\ 
    & + (1-p) q_{2} \bar{V}_{\alpha,n-1}^{N} ((\bar{v}_{2}+1, \infty, 0)) \nonumber\\  
    & + (1-p)(1-q_{2}) \bar{V}_{\alpha,n-1}^{N}((v_{1}+1, \bar{v}_{2}+1, 0)) \nonumber\\ 
    &+ p (1-q_{2}) \bar{V}_{\alpha,n-1}^{N}((v_{1}+1, \bar{v}_{2}+1,1))
\end{align}
Now, from the assumption of the induction stage $n-1$,
\begin{align}\label{eq:40}
 \pdv{\bar{V}_{\alpha,n-1}^{N}((v_{1}\!+\!1, v_{2}\!+\!1,1))^{-}}{v_{1}} \!\leq\! \pdv{\bar{V}_{\alpha,n-1}^{N}((v_{1}\!+\!1, \bar{v}_{2}\!+\!1, 1))^{-}}{v_{1}}
\end{align}
Thus, from (\ref{eq:38}), (\ref{eq:39}) and (\ref{eq:40}), we say that for the first case,
\begin{align}\label{eq:41}
    \pdv{\bar{V}_{\alpha,n}^{N}({s'})^{-}}{v_{1}}\leq\pdv{\bar{V}_{\alpha,n}^{N}(\bar{s}_{3})^{-}}{v_{1}}
\end{align}
Similarly, we show that for the first case, 
\begin{align}\label{eq:42}
    \pdv{\bar{V}_{\alpha,n}^{N}(\bar{s}_{3})^{-}}{v_{1}}\leq\left(\frac{1-q_{1}}{1-q_{2}}\right)\pdv{\bar{V}_{\alpha,n}^{N}({s'})^{-}}{v_{1}}
\end{align}
Further, we show the inequalities when the optimal actions for both states $s'$, $\bar{s}_{3}$ is $a=1$, at the induction stage $n$. 

Now, consider the case when the optimal action for $s'$ is $a=0$, however, the optimal action for $\bar{s}_{3}$ is $a=1$. Then,
\begin{align}\label{eq:43}
     \bar{V}_{\alpha,n}^{N}(s') =& q_{2} (v_{2}+1) + (1-q_{2}) (v_{1}+1) \nonumber\\
     & + p q_{2} \bar{V}_{\alpha,n-1}^{N} ((v_{2}+1, \infty, 1)) \nonumber\\ 
     & +  (1-p) q_{2} \bar{V}_{\alpha,n-1}^{N}((v_{2}+1,\infty,0)) \nonumber\\ 
     & + (1-p)(1-q_{2}) \bar{V}_{\alpha,n-1}^{N}((v_{1}+1, v_{2}+1, 0)) \nonumber\\
     & + p (1-q_{2}) \bar{V}_{\alpha,n-1}^{N}((v_{1}+1, v_{2}+1, 1))
\end{align}
Similarly,
\begin{align}\label{eq:44}
    \bar{V}_{\alpha,n}^{N}(\bar{s}_{3}) = & q_{1} + (1-q_{1}) (v_{1}+1) + p q_{1} \bar{V}_{\alpha,n-1}^{N} ((1, \infty, 1))\nonumber\\
    & + (1-p) q_{1} \bar{V}_{\alpha,n-1}^{N}((1,\infty,0)) \nonumber\\
    & + p (1-q_{1}) \bar{V}_{\alpha,n-1}^{N}((v_{1}+1, 1, 1)) \nonumber\\
    & + (1-p)(1-q_{1}) \bar{V}_{\alpha,n-1}^{N}((v_{1}+1, 1, 0))
\end{align}
Now, from the assumption on the induction step $n-1$, 
\begin{align}
    \!\pdv{\bar{V}_{\alpha,n-1}^{N}((v_{1}\!+\!1,\!v_{2}\!+\!1,\!b))^{-}}{v_{1}} \!\!\leq\!\! \left(\!\frac{1\!-\!q_{1}}{1\!-\!q_{2}}\!\right) \!\pdv{\bar{V}_{\alpha,n-1}^{N}((v_{1}\!+\!1,\!1,\!b))^{-}}{v_{1}}
\end{align}
Thus, from (\ref{eq:43}) and (\ref{eq:44}), we say that for this case at the induction stage $n$,
\begin{align}\label{eq:46}
    \pdv{\bar{V}_{\alpha,n}^{N}({s'})^{-}}{v_{1}}\leq\pdv{\bar{V}_{\alpha,n}^{N}(\bar{s}_{3})^{-}}{v_{1}}
\end{align}

Now, combining all the above mentioned three cases, we say that (\ref{eq:46}) is always true. By similar techniques, we show the the proofs of the following inequalities,
\begin{align}
         \pdv{\bar{V}_{\alpha,n}^{N}(\bar{s}_{3})^{-}}{v_{1}} & \leq \left(\frac{1-q_{1}} {1-q_{2}}\right)  \pdv{\bar{V}_{\alpha,n}^{N}(s')^{-}}{v_{1}}\label{eq:47}  \\  
         \pdv{\bar{V}_{\alpha,n}^{N}((v_{1},v_{2},0))^{-}}{v_{1}} & \leq  \pdv{\bar{V}_{\alpha,n}^{N}((v_{1},\bar{v}_{2},0))^{-}}{v_{1}}\label{eq:48} \\   
         \pdv{\bar{V}_{\alpha,n}^{N}((v_{1},\bar{v}_{2},0))^{-}}{v_{1}} & \leq \left(\frac{1-q_{1}}{1-q_{2}}\right) \pdv{\bar{V}_{\alpha,n}^{N}((v_{1},v_{2},0))^{-}}{v_{1}}\label{eq:49}
\end{align}

For the case $(N-1)<v_{1}\leq N$, we can show similar inequalities as in (\ref{eq:46})-(\ref{eq:49}). Thus, for $n\geq 1$, $N\geq 1$, $1 \leq v_{1}\leq N$, (\ref{eq:46}-(\ref{eq:49}) holds true. Now, we want to find the limiting values for (\ref{eq:46})-(\ref{eq:49}). Next, we want to show the following,
\begin{align}
    \lim_{n\rightarrow\infty} \pdv{\bar{V}_{\alpha,n}^{N}(\bar{s}_{3})^{-}}{v_{1}} = \pdv{\bar{V}_{\alpha}^{N}(\bar{s}_{3})^{-}}{v_{1}}
\end{align}
and the rest of the inequalities follow similarly. Note that, 
\begin{align}
    \bar{V}_{\alpha,n}^{N}(\check{s}) \leq \frac{N}{1-\alpha}, \quad \check{s}\in{\bar{\mathcal{S}}_{N}}
\end{align}
Thus, from \cite[Prop.~1]{ross2014introduction},  we say that,
\begin{align}
\left|\bar{V}_{\alpha}^{N}(\check{s})-\bar{V}_{\alpha,n}^{N}(\check{s})\right| \leq \frac{\alpha^{n+1} N}{1-\alpha}, \quad n\geq 1, \ \check{s}\in{\bar{\mathcal{S}}_{N}}
\end{align}
Thus, the sequence $\{\bar{V}_{\alpha,n}^{N}(\check{s})\}_{n=1}^{\infty}$ is uniformly convergent on $\bar{\mathcal{S}}_{N}$. Therefore,
\begin{align}
     &\lim_{n\rightarrow\infty} \pdv{\bar{V}_{\alpha,n}^{N}(\bar{s}_{3})^{-}}{v_{1}} \nonumber\\ &=\lim_{n\rightarrow \infty} \lim_{h\rightarrow 0+} \frac{\bar{V}_{\alpha,n}^{N}((v_{1},\bar{v}_{2},1)) - \bar{V}_{\alpha,n}^{N}((v_{1}\!-\!h,\bar{v}_{2},1))}{h} \\ 
     &=\lim_{h\rightarrow 0+}\lim_{n\rightarrow\infty}  \frac{\bar{V}_{\alpha,n}^{N}((v_{1},\bar{v}_{2},1)) - \bar{V}_{\alpha,n}^{N}((v_{1}\!-\!h,\bar{v}_{2},1))}{h} \label{eq:53}  \\ 
     & = \pdv{\bar{V}_{\alpha}^{N}(\bar{s}_{3})^{-}}{v_{1}} \label{eq:54}
\end{align}
where (\ref{eq:53}) follows from the Moore-Osgood theorem. 

For the next part of the proof, we write $\bar{V}_{\alpha}^{N}((v_{1}, v_{2}, 1)) = g(v_{1})$, only a function of the age of the user, $v_{1}$. Let us define, $\epsilon_{1}>0$ and $\epsilon_{2}>0$, which are small enough such that $[1+\epsilon_{1}, N-\epsilon_{2}]$ is a non-empty interval. From Lemma~\ref{lemma:6}, we know that $g(v_{1})$ is a concave function of $v_{1}$. Thus, $ \pdv{g(v_{1})^{-}}{v_{1}}$ is monotonically decreasing and $g$ is continuous on $(1, N)$ \cite{bertsekas2009convex}. Combining these two statements we say that $\pdv{g(v_{1})^{-}}{v_{1}}$ is finite on $[1+\epsilon_{1}, N- \epsilon_{2}]$, thus $\pdv{g(v_{1})^{-}}{v_{1}}$ is bounded by $\pdv{g(1+\epsilon_{1})^{-}}{v_{1}}$. Consider $x_{1}, x_{2} \in{[1+\epsilon_{1}, N- \epsilon_{2}]}$ such that $x_{1}<x_{2}$. Take $h>0$, small enough such that $1<x_{1}-h$ and $x_{1}<x_{2}-h$. Now, from the convexity of $g(v_{1})$,
\begin{align}
   \frac{g(x_{1})\!-\!g(x_{1}\!-\!h)}{h} \!\geq\! \frac{g(x_{2}) \!-\! g(x_{1})}{x_{2}\!-\!x_{1}} \!\geq\! \frac{g(x_{2})\!-\! g(x_{2}\!-\!h)}{h}
\end{align}
thus,
\begin{align}
    \pdv{g(v_{1})^{-}}{v_{1}}\Bigg|_{x_{1}} \geq \frac{g(x_{2}) - g(x_{1})}{x_{2}-x_{1}} \geq \pdv{g(v_{1})^{-}}{v_{1}}\Bigg|_{x_{2}}\label{eq:56}
\end{align}

Consider a partition $P$ on $[1+\epsilon_{1}, N- \epsilon_{2}]$, $x_{0}=1+\epsilon_{1}<x_{1}<x_{2}<\cdots<x_{n}=N-\epsilon_{2}$.
Then, from (\ref{eq:56}), 
\begin{align}
    & \sum_{i=1}^{n} \pdv{g(v_{1})^{-}}{v_{1}}\Bigg|_{x_{i-1}} (x_{i}- x_{i-1}) \geq \sum_{i=1}^{n} g(x_{i}) - g(x_{i-1}) \nonumber\\ 
    & \quad \geq  \sum_{i=1}^{n} \pdv{g(v_{1})^{-}}{v_{1}}\Bigg|_{x_{i}} (x_{i}- x_{i-1})
\end{align}
Thus,
\begin{align}\label{eq:58}
      U\left(\pdv{g(v_{1})^{-}}{v_{1}},P\right) & \geq  g(N-\epsilon_{2}) - g(1+\epsilon_{1}) \nonumber\\ 
      & \geq L\left(\pdv{g(v_{1})^{-}}{v_{1}},P\right)
\end{align}
where $ U\left(\pdv{g(v_{1})^{-}}{v_{1}},P\right)$ and $L\left(\pdv{g(v_{1})^{-}}{v_{1}},P\right)$ are the upper and the lower Riemann sums for the partition $P$. Again, consider the same partition and assume that $\max_{1\leq i\leq n}{(x_{i}-x_{i-1})} \leq \epsilon_{3}$. As $\pdv{g(v_{1})^{-}}{v_{1}}$ is monotone, we say that,
\begin{align}
    & \left|U\left(\pdv{g(v_{1})^{-}}{v_{1}},P\right) - L\left(\pdv{g(v_{1})^{-}}{v_{1}},P\right)  \right| \nonumber\\ 
    & \leq\left|\sum_{i=1}^{n} \left(\pdv{g(v_{1})^{-}}{v_{1}}\bigg|_{x_{i-1}} - \pdv{g(v_{1})^{-}}{v_{1}}\bigg|_{x_{i}}\right) (x_{i}-x_{i-1})\right| \\  
    & \leq \epsilon_{3} \left|\left(\pdv{g(v_{1})^{-}}{v_{1}}\bigg|_{N-\epsilon_{2}} - \pdv{g(v_{1})^{-}}{v_{1}}\bigg|_{1+\epsilon_{1}}\right)\right|\label{eq:59} 
\end{align}
As $\pdv{g(v_{1})^{-}}{v_{1}}$ is also bounded on $[1+\epsilon_{1}, N- \epsilon_{2}]$, from (\ref{eq:59}), there exists a sequence of partitions $\{P_{n}\}_{n=1}^{\infty}$, such that,
\begin{align}
    \lim_{n\rightarrow\infty} \left|L\left(\pdv{g(v_{1})^{-}}{v_{1}},P_{n}\right) - U\left(\pdv{g(v_{1})^{-}}{v_{1}},P_{n}\right)\right|=0
\end{align}
Now, from (\ref{eq:58}), we say that,
\begin{align}
    \lim_{n\rightarrow\infty} U\left(\pdv{g(v_{1})^{-}}{v_{1}},P_{n}\right) & =  \lim_{n\rightarrow\infty}  L\left(\pdv{g(v_{1})^{-}}{v_{1}},P_{n}\right) \\ 
    & = \int_{1+\epsilon_{1}}^{N-\epsilon_{2}} \pdv{g(v_{1})^{-}}{v_{1}} dv_{1} \\
    & = g(N-\epsilon_{2}) - g(1+\epsilon_{1}) \label{eq:61}
\end{align}
Now, as $g$ is a continuous function of $v_{1}$, we first take $\epsilon_{1}\rightarrow 0$, then take $\epsilon_{2}\rightarrow 0$, and finally, we get,
\begin{align}
\int_{1}^{N} \pdv{g(v_{1})^{-}}{v_{1}} dv_{1} = g(N) - g(1)
\end{align}
From (\ref{eq:46}), (\ref{eq:54}) and with similar argument as above, 
\begin{align}
 \int_{v_{1}}^{v_{1}+y} \pdv{\bar{V}_{\alpha}^{N}(x,v_{2},1)^{-}}{v_{1}} dx \leq  \int_{v_{1}}^{v_{1}+y} \pdv{\bar{V}_{\alpha}^{N}(x,\bar{v}_{2},1)^{-}}{v_{1}} 
\end{align}
Thus,
\begin{align}
     \bar{V}_{\alpha}^{N}((v_{1}&+y,v_{2},b))-\bar{V}_{\alpha}^{N}((v_{1}, {v}_{2}, b )) \nonumber\\ & \leq \bar{V}_{\alpha}^{N}((v_{1}+y, \bar{v}_{2},b)) - \bar{V}_{\alpha}^{N}((v_{1},\bar{v}_{2},b))
\end{align}
In a similar fashion, we show all the inequalities considered in the statement of this lemma for $\bar{V}_{\alpha}^{N}(\cdot)$. Now, if $v_{1}, v_{1}+y, v_{2}, \bar{v}_{2} \in{\mathbb{N}}$, $\bar{V}_{\alpha}^{N}(\cdot)$ and $V_{\alpha}^{N}(.)$ are the same, this concludes the proof of this lemma.
\end{Proof}

Next, we study another structure of the optimal policy.
\begin{theorem}\label{th:2}
   If $v_{1}\geq \frac{q_{2} v_{2}}{q_{2}-q_{1}}$, and if the optimal action for the state $(v_{1}, v_{2}, 1)$ is to choose action $0$, then for any non-negative $x$, the optimal action for the state $(v_{1}+x, v_{2}, 1)$ is also to choose action $0$.
\end{theorem}

\begin{Proof}
Due to space restrictions, we only provide a sketch of the proof here: For $N\geq 1$, we use induction, Lemma~\ref{lemma:7} and arguments similar to those in the proof for Lemma~\ref{lemma:4} to prove the statement of this theorem for $\Delta_{N}$. Then, we use Lemma~\ref{lemma:5} to conclude the proof of this theorem. 
\end{Proof}

For an MDP, if all the stationary policies induce a uni-chain, then we say that the MDP is uni-chain. In the next theorem we show that the $\Delta$ is an uni-chain MDP.

\begin{theorem}
    $\Delta$ is an uni-chain MDP.
\end{theorem}

\begin{Proof}
   We only provide a sketch of the proof here: Note that, any policy which transmits when a fresh update packet is available can reach the state $(1,\infty,1)$ and from this state every other state is reachable. Thus, $\Delta$ is a uni-chain MDP.
\end{Proof}

As the MDP $\Delta$ is a uni-chain, next, we use the well-known relative value iteration algorithm (RVIA) \cite{puterman2014markov} to find the optimal stationary deterministic algorithm for (\ref{eq:5}). For $N\geq 1$, RVIA for $\Delta_{N}$ and for all the states $\hat{s}\in{\mathcal{S}_{N}}$ is,
\begin{align}\label{eq:87}
    V_{n}^{N}(\hat{s}) =  \min_{a\in{\{0,1,2\}}}& \Big\{C(\hat{s},a) + \sum_{\tilde{s}\in{\mathcal{N}}} P_{a}(\hat{s},\tilde{s}) V_{n-1}^{N}(\tilde{s}) \nonumber\\ 
    & \ - V_{n-1}^{N}((1,1,1))\Big\}
\end{align}
where $V_{n}^{N}(\hat{s})$ is the value function corresponding to the average cost optimality criterion for $\Delta_{N}$. We initialize $V_{0}^{N}(s) = 0$, for all $s\in{\mathcal{S}_{N}}$. Now, from Lemma~\ref{lemma:5}, for large $N$, we obtain the optimal stationary deterministic policy for (\ref{eq:5}). We use the structural properties of the optimal policy, i.e., Lemma~\ref{lemma:4} and Theorem~\ref{th:2} to reduce the  complexity of RVIA in (\ref{eq:87}).

\bibliographystyle{unsrt}
\bibliography{references}
\end{document}